\documentclass[twocolumn]{aastex63}
\hypersetup{linkcolor=red,citecolor=blue,filecolor=cyan,urlcolor=black}

\newcommand{\OIII}{O\,{\scriptsize III}}

\newcommand{\CII}{C\,{\scriptsize II}}

\newcommand{\NIII}{N\,{\scriptsize III}}
\newcommand{\CIII}{C\,{\scriptsize III}}

\newcommand{\NV}{N\,{\scriptsize V}}
\newcommand{\CIV}{C\,{\scriptsize IV}}
\newcommand{\HeII}{He\,{\scriptsize II}}

\newcommand{\hst}{\textit{HST}}
\newcommand{\jwst}{\textit{JWST}}

\newcommand{\grizli}{\texttt{Grizli}}

\graphicspath{{./}{figures/}}

\shorttitle{\textit{JWST}/NIRISS Spectroscopy of $z\geqslant7$ Galaxies}
\shortauthors{Roberts-Borsani et al.}

\begin{document}



\title{Early Results from GLASS-JWST. I: Confirmation of Lensed $z\geqslant7$ Lyman-Break Galaxies Behind the Abell 2744 Cluster With NIRISS}

\correspondingauthor{Guido Roberts-Borsani}
\email{guidorb@astro.ucla.edu}

\author[0000-0002-4140-1367]{Guido Roberts-Borsani}
\affiliation{Department of Physics and Astronomy, University of California, Los Angeles, 430 Portola Plaza, Los Angeles, CA 90095, USA}

\author[0000-0002-8512-1404]{Takahiro Morishita}
\affiliation{Infrared Processing and Analysis Center, Caltech, 1200 E. California Blvd., Pasadena, CA 91125, USA}

\author[0000-0002-8460-0390]{Tommaso Treu}
\affiliation{Department of Physics and Astronomy, University of California, Los Angeles, 430 Portola Plaza, Los Angeles, CA 90095, USA}

\author[0000-0003-2680-005X]{Gabriel Brammer}
\affiliation{Cosmic Dawn Center (DAWN), Denmark}
\affiliation{Niels Bohr Institute, University of Copenhagen, Jagtvej 128, DK-2200 Copenhagen N, Denmark}

\author[0000-0002-6338-7295]{Victoria Strait}
\affiliation{Cosmic Dawn Center (DAWN), Denmark}
\affiliation{Niels Bohr Institute, University of Copenhagen, Jagtvej 128, DK-2200 Copenhagen N, Denmark}

\author[0000-0002-9373-3865]{Xin Wang}
\affiliation{Infrared Processing and Analysis Center, Caltech, 1200 E. California Blvd., Pasadena, CA 91125, USA}

\author[0000-0001-5984-0395]{Marusa Bradac}
\affiliation{University of Ljubljana, Department of Mathematics and Physics, Jadranska ulica 19, SI-1000 Ljubljana, Slovenia}
\affiliation{Department of Physics and Astronomy, University of California Davis, 1 Shields Avenue, Davis, CA 95616, USA}


\author[0000-0003-3108-9039]{Ana Acebron}
\affiliation{Dipartimento di Fisica, Università degli Studi di Milano, Via Celoria 16, I-20133 Milano, Italy}
\affiliation{INAF - IASF Milano, via A. Corti 12, I-20133 Milano, Italy}

\author[0000-0003-1383-9414]{Pietro Bergamini}
\affiliation{Dipartimento di Fisica, Università degli Studi di Milano, Via Celoria 16, I-20133 Milano, Italy}
\affiliation{INAF - OAS, Osservatorio di Astrofisica e Scienza dello Spazio di Bologna, via Gobetti 93/3, I-40129 Bologna, Italy}

\author[0000-0003-4109-304X]{Kristan Boyett}
\affiliation{School of Physics, University of Melbourne, Parkville 3010, VIC, Australia}
\affiliation{ARC Centre of Excellence for All Sky Astrophysics in 3 Dimensions (ASTRO 3D), Australia}

\author[0000-0003-2536-1614]{Antonello Calabr\'o}
\affiliation{INAF Osservatorio Astronomico di Roma, Via Frascati 33, 00078 Monteporzio Catone, Rome, Italy}

\author[0000-0001-9875-8263]{Marco Castellano}
\affiliation{INAF Osservatorio Astronomico di Roma, Via Frascati 33, 00078 Monteporzio Catone, Rome, Italy}

\author[0000-0003-3820-2823]{Adriano Fontana}
\affiliation{INAF Osservatorio Astronomico di Roma, Via Frascati 33, 00078 Monteporzio Catone, Rome, Italy}

\author[0000-0002-3254-9044]{Karl Glazebrook}
\affiliation{Centre for Astrophysics and Supercomputing, Swinburne University of Technology, PO Box 218, Hawthorn, VIC 3122, Australia}

\author[0000-0002-5926-7143]{Claudio Grillo}
\affiliation{Dipartimento di Fisica, Università degli Studi di Milano, Via Celoria 16, I-20133 Milano, Italy}
\affiliation{INAF - IASF Milano, via A. Corti 12, I-20133 Milano, Italy}

\author[0000-0002-6586-4446]{Alaina Henry}
\affiliation{Space Telescope Science Institute, 3700 San Martin Drive, Baltimore MD, 21218} 
\affiliation{Center for Astrophysical Sciences, Department of Physics and Astronomy, Johns Hopkins University, Baltimore, MD, 21218}

\author[0000-0001-5860-3419]{Tucker Jones}
\affiliation{Department of Physics and Astronomy, University of California Davis, 1 Shields Avenue, Davis, CA 95616, USA}

\author[0000-0001-6919-1237]{Matthew Malkan}
\affiliation{Department of Physics and Astronomy, University of California, Los Angeles, 430 Portola Plaza, Los Angeles, CA 90095, USA}

\author[0000-0001-9002-3502]{Danilo Marchesini}
\affiliation{Department of Physics and Astronomy, Tufts University, 574 Boston Ave., Medford, MA 02155, USA}

\author[0000-0002-9572-7813]{Sara Mascia}
\affiliation{INAF Osservatorio Astronomico di Roma, Via Frascati 33, 00078 Monteporzio Catone, Rome, Italy}

\author[0000-0002-3407-1785]{Charlotte Mason}
\affiliation{Cosmic Dawn Center (DAWN), Denmark}
\affiliation{Niels Bohr Institute, University of Copenhagen, Jagtvej 128, DK-2200 Copenhagen N, Denmark}

\author[0000-0001-9261-7849]{Amata Mercurio}
\affiliation{INAF -- Osservatorio Astronomico di Capodimonte, Via Moiariello 16, I-80131 Napoli, Italy}

\author[0000-0001-6870-8900]{Emiliano Merlin}
\affiliation{INAF Osservatorio Astronomico di Roma, Via Frascati 33, 00078 Monteporzio Catone, Rome, Italy}

\author[0000-0003-2804-0648 ]{Themiya Nanayakkara}
\affiliation{Centre for Astrophysics and Supercomputing, Swinburne University of Technology, PO Box 218, Hawthorn, VIC 3122, Australia}

\author[0000-0001-8940-6768 ]{Laura Pentericci}
\affiliation{INAF Osservatorio Astronomico di Roma, Via Frascati 33, 00078 Monteporzio Catone, Rome, Italy}

\author[0000-0002-6813-0632]{Piero Rosati}
\affiliation{Dipartimento di Fisica e Scienze della Terra, Università degli Studi di Ferrara, Via Saragat 1, I-44122 Ferrara, Italy}
\affiliation{INAF - OAS, Osservatorio di Astrofisica e Scienza dello Spazio di Bologna, via Gobetti 93/3, I-40129 Bologna, Italy}

\author[0000-0002-9334-8705]{Paola Santini}
\affiliation{INAF Osservatorio Astronomico di Roma, Via Frascati 33, 00078 Monteporzio Catone, Rome, Italy}

\author[0000-0002-9136-8876]{Claudia Scarlata}\affiliation{School of Physics and Astronomy, University of Minnesota, Minneapolis, MN, 55455, USA}

\author[0000-0001-9391-305X]{Michele Trenti}
\affiliation{School of Physics, University of Melbourne, Parkville 3010, VIC, Australia}
\affiliation{ARC Centre of Excellence for All Sky Astrophysics in 3 Dimensions (ASTRO 3D), Australia}

\author[0000-0002-5057-135X]{Eros Vanzella}
\affiliation{INAF -- OAS, Osservatorio di Astrofisica e Scienza dello Spazio di Bologna, via Gobetti 93/3, I-40129 Bologna, Italy}

\author[0000-0003-0980-1499]{Benedetta Vulcani}
\affiliation{INAF Osservatorio Astronomico di Padova, vicolo dell'Osservatorio 5, 35122 Padova, Italy}

\author[0000-0002-4201-7367]{Chris Willott}
\affiliation{NRC Herzberg, 5071 West Saanich Rd, Victoria, BC V9E 2E7, Canada}

\begin{abstract}
We present the first search for $z\geqslant7$, continuum-confirmed sources with NIRISS/WFS spectroscopy over the Abell 2744 Frontier Fields cluster, as part of the GLASS-JWST ERS survey. With $\sim15$ hrs of pre-imaging and multi-angle grism exposures in the F115W, F150W, and F200W filters, we describe the general data handling (i.e., reduction, cleaning, modeling, and extraction processes) and analysis for the GLASS-JWST survey. We showcase the power of \textit{JWST} to peer deep into reionization, when most intergalactic hydrogen is neutral, by confirming two galaxies at $z=8.04\pm0.15$ and $z=7.90\pm0.13$ by means of their Lyman breaks. Fainter continuum spectra are observed in both the F150W and F200W bands, indicative of blue ($-1.69$ and $-1.33$) UV slopes and moderately-bright absolute magnitudes ($-20.37$ and $-19.68$ mag). We do not detect strong Ly$\alpha$ in either galaxy, but do observe tentative ($\sim2.7-3.8\sigma$) \HeII$\lambda$1640 \AA, \OIII]$\lambda\lambda$1661,1666 \AA, and \NIII]$\lambda\lambda$1747,1749 \AA\ line emission in one, suggestive of low metallicity, star-forming systems with possible non-thermal contributions. These novel observations provide a first look at the extraordinary potential of \textit{JWST}/NIRISS for confirming representative samples of bright $z\geqslant7$ sources in the absence of strong emission lines, and gain unprecedented insight into their contributions towards cosmic reionization.
\end{abstract}


\section{Introduction}
The arrival of the \textit{James Webb} Space Telescope (\textit{JWST}) heralds a new era for the study of early galaxy evolution. The extension of imaging capabilities to higher angular resolution and into the infrared (IR) regime will allow for the identification of Lyman break galaxies (LBGs) well beyond the current redshift horizon of $z\sim10$ \citep[e.g.,][]{castellano22}, set by the \textit{Hubble} Space Telescope (\textit{HST}). Furthermore, \textit{JWST}'s unprecedented spectroscopic capabilities at wavelengths of $\lambda > 1~\mu$m will prove to be a game-changer for the characterization of their ionizing capabilities, underlying gas conditions, and stellar populations through emission line (e.g., Ly$\alpha$, \CIII]$\lambda\lambda$1907,1909 \AA, \HeII$\lambda$1640 \AA, H$\beta\lambda$4861 \AA, [\OIII]$\lambda$5007 \AA, and H$\alpha\lambda$6563 \AA) and direct continuum measurements (e.g., the 4000 \AA\ Balmer Break and robust UV spectral slopes), respectively.


While \hst\ and ground-based surveys have seen remarkable success in the identification of $z>7$ galaxy candidates \citep{Castellano2010b,schmidt14,bouwens15,oesch18,morishita18,bowler20,strait21,finkelstein22,rb22} from deep fields (e.g., the \hst\ Ultra Deep Field 2012, the Cosmic Assembly Near-Infrared Deep Extragalactic Legacy Survey, and UltraVISTA; \citealt{ellis13,grogin11, mccracken12}, respectively), lensing clusters (e.g., the Cluster Lensing And Supernova survey with Hubble, the Frontier Fields, and the Reionization Lensing Cluster Survey; \citealt{postman12,lotz17,coe19}, respectively) and pure-parallel data sets (e.g., the Brightest of Reionizing Galaxies survey; \citealt{trenti11,morishita21}), spectroscopic confirmation of those candidates - a prerequisite for accurate characterizations of their underlying properties - has remained elusive.

Confirmations via Ly$\alpha$ (which is strongly affected by the surrounding intergalactic medium; \citealt{treu13,mason19}) and other rest-frame UV lines in the near-infrared (NIR) with e.g., Keck and the VLT has seen limited success and been found predominantly in the most luminous sources (e.g., \citealt{Vanzella2011,finkelstein15,oesch15,zitrin15,rb16,laporte17b,stark17,mainali18,mason19,hoag19,endsley21,laporte21,rb22_mos}). ALMA has provided an alternate avenue for spectroscopic confirmation and characterization in the far-infrared (FIR), with an increasing number of [\OIII] 88 $\mu$m, [\CII] 158 $\mu$m, and/or dust continuum detections \citep[e.g., ][]{laporte17,hashimoto18,tamura19,bouwens21_rebels}. However, in both cases high redshift observations are strongly affected by the intervening Earth atmosphere, highlighting the need for space-based observations. 


Importantly, one major limitation of $z\geqslant7$ galaxy studies is the near exclusivity of redshift determinations through bright emission lines, i.e., Ly$\alpha$ and [\OIII] 88 $\mu$m. However, as one enters the reionization era, Ly$\alpha$ becomes a biased tracer of exceptional systems and environments due to its attenuation by a neutral IGM. FIR lines, in contrast, have only been observed in the most luminous, dusty, and star-forming systems \citep[e.g., ][]{laporte17,bouwens21_rebels}. As such, a large and key population of LBGs that generally do not show Ly$\alpha$ emission \citep{giavalisco02,shapley11} currently lack spectroscopic confirmations, potentially skewing our understanding of galaxy properties and our interpretation of the reionization process and its main drivers. Grism spectroscopy with \hst\ has shown promise for high redshift studies \citep{watson15,schmidt16,treu15,oesch16,hoag18}, however \hst's instruments lack the sensitivity required for systematic confirmations of $z\geqslant7$ populations. \jwst's unrivaled sensitivity and spectroscopic capabilities thus offer an unprecedented opportunity to confirm redshifts for this missing sample through continuum measurements (independently of Ly$\alpha$), paving the way for subsequent characterization of ISM conditions, stellar population properties, and impact on the surrounding IGM for a representative population of galaxies.

Here we present the first search for $z\geqslant7$ continuum sources behind the Abell 2744 Frontier Field galaxy cluster as part of the GLASS-JWST program \citep[ERS 1324, PI Treu;][]{treu22}, one of the very first and deepest extragalactic data sets of the Early Release Science (ERS) campaign.
The search, combined with the lensing magnification afforded by the foreground cluster, showcases the potential of \textit{JWST} to peer deep into the reionization era independently of Ly$\alpha$. The Letter is structured as follows. In Section~\ref{sec:data} we describe our reduction of the data, contamination modeling, and modeling and extraction of grism spectra. In Section~\ref{sec:results} we describe our target selection and showcase our results. We provide a summary and conclusions in Section~\ref{sec:conclusions}.
This Letter serves as a reference for the data reduction and general modeling procedures employed by all the NIRISS-based papers in this Focus Issue. Where relevant, we assume \textit{H}$_{0}=$70 km/s/Mpc, $\Omega_{m}=$0.3, and $\Omega_{\wedge}=$0.7. All magnitudes are in the AB system \citep{oke83}.

\section{The \textit{JWST} GLASS-ERS Data Set}
\label{sec:data}
\subsection{Data Reduction}
We focus on the GLASS-JWST-ERS NIRISS observations of the central regions of the Abell 2744 cluster, obtained on June 28-29 2022. The field was observed with $\sim15$ hrs of wide field slitless spectroscopy (WFSS; \citealt{willot22}) at $R\approx150$ spectral resolution (at two orthogonal angles) and moderately deep ($\sim28.6-28.9$ AB at $5\sigma$; \citealt{treu22}) pre-imaging in three different filters (F115W, F150W, F200W). The wavelength range afforded by the three filters allows for the identification of a large variety of spectral features across a number of object types and redshifts, while the choice of two orthogonal grisms (GR150C and GR150R) facilitates the disentanglement of spectra in crowded environments (see \citealt{treu22} for details on the observing strategy).

We reduce the entire data set using the latest set of available reference files (``jwst\_0916.pmap'', which includes in-flight calibrations) and the latest version the Grism Redshift \& Line \citep[\texttt{Grizli\footnote{\url{https://github.com/gbrammer/grizli}}};][]{brammer21} analysis software, which incorporates the majority of \texttt{Python} routines from the STScI data reduction pipeline as well as custom routines for additional improvements (e.g., background subtraction, image alignment, and drizzling). Starting with the available count-rate files generated by the \texttt{Detector1} STScI pipeline (which applies detector-level corrections such as the identification of bad pixels and cosmic rays, subtraction of dark current, and ramp fitting), we run \grizli's \texttt{preprocessing} pipeline
which performs WCS registration and astrometric alignment, flat-fielding, sky background subtraction, and pixel drizzling to provide fully reduced individual exposures and mosaics for both the pre-imaging and WFSS data sets. We align all our images to the LegacySurveys DR9 \citep{dey19} astrometry, in order to match the astrometric reference frame used for the ALMA Lensing Cluster Survey's data reduction of \hst\ data \citep{kokorev22}. For image drizzling, we produce two mosaics at 30 milli-arcseconds (mas) and at 60 mas, of which the former can be used for resolved studies and the latter for higher S/N studies of point-like sources. We adopt the latter throughout this Letter. We calculate the 5$\sigma$ limiting magnitudes for each of the stacked (and drizzled) pre-images: we carefully place 10 circular, $r=0.1''$ apertures at representative locations across the images - ensuring these remain free of signal or rare image artifacts - and measure the median standard deviation from those, before scaling and converting to the desired units. We find our stacked images reach depths of 28.2-28.5 AB (or 29.3 AB for a stacked IR image, see below), consistent with our estimates in \citet{treu22} adopting identically-sized apertures.


\subsection{Removal of Contaminating Sources}
\label{subsec:contam}
One of the most significant challenges of slitless spectroscopy in clustered environments is to accurately remove overlapping spectra of nearby sources \citep[e.g.,][]{treu15,oesch16,schmidt16}, given the tendency for contamination to cause confusion in redshift estimations and physical interpretation of a source's spectrum. Considering the large density of sources in our field, such challenges are especially relevant. To identify the position of sources in the  $\sim2.1'\times2.1'$ field-of-view (FOV), we begin by constructing a flux-weighted IR stack from the pre-imaging mosaics, which is used to generate a catalog of sources and associated segmentation map. For each detected source brighter than $m_{\rm AB}=$28, \grizli\ then fits and refines a 7th order polynomial to the associated spectrum in each individual grism exposure, thereby creating a map with which to identify contaminating pixels around or nearby objects of interest. For F115W spectra (where spatial offsets between the target position in direct imaging and the spectral trace are negligible), we find the contamination model is able to adequately subtract the majority of nearby contaminants, however this becomes more challenging as a function of wavelength where offsets of the spectral trace relative to the source become larger. Improvements and imminent updates to existing reference files will further improve wavelength calibrations, spectral trace offset measurements, and flux calibrations for more precise measurements. Considering the above, all pixels used for extraction and modeling purposes are therefore weighted by $w_{\rm pix}$ according to a combination of their contamination level and flux uncertainty (defined by the error array in the flux-calibrated count-rate images). We define the weights as $w_{\rm pix}=\sigma_{\rm pix}^{-2}\cdot\exp(-f_{\rm c} \cdot |f_{\lambda,\rm cont,pix}| / \sqrt{\sigma_{\rm pix}^{2}})$, where $f_{\lambda,\rm cont,pix}$ is the contamination model flux, $f_{\rm c}$ is a constant down-weighting factor set to 0.2, and $\sigma_{\rm pix}$ is the flux noise for a pixel given by the error array.



\subsection{Modeling \& Extraction of Grism Spectra}
For a specified target, all individual, ``cleaned'' (i.e., weighted according to the equation described above) 2D grism exposures provided to \grizli\ are modelled simultaneously, allowing for independent noise handling and the avoidance of additional complications resulting from stacking (e.g., morphological broadening and smearing of spectral traces from asymmetric pixel sizes, orthogonal dispersion directions, and varying trace offsets between grism setups), while retaining a collective S/N ratio equivalent to that of a stacked mosaic. Using constraints from the collective sample of exposures, a 2D galaxy model is generated via a linear combination of galaxy templates and dispersed onto the plane of the sky according to the location and morphological structure of the source as measured from the pre-imaging exposures. To  generate the model, we make use of variety of galaxy templates (e.g., intermediate age SEDs with moderate 4000 \AA\ breaks, older simple stellar population models to account for low-level absorption features, post-starburst galaxies from the UltraVISTA survey, and low-metallicity LBGs) incorporating continuum and emission line contributions, to be able to flexibly fit a large variety of spectra according to an allowed redshift range. 

Given the common astrometric reference frame, in this study we also include for added constraints the full assortment of available \hst\ photometry from the Frontier Fields, derived from the drizzled mosaics of \citet{kokorev22} that we resampled to 60 mas pixel size to match our NIRISS mosaics. Specifically, we include photometry from \hst/WFC3-UVIS (F336W), \hst/ACS (F435W, F606W, F814W) and \hst/WFC3 (F105W, F125W, F140W, and F160W), which serve as valuable additional constraints to determine the location of the Lyman break. As an added check, for sources of interest, we also visually inspect each filter-grism stack and exclude combinations where we deem the spectra to be significantly impacted by lingering contamination (e.g., if they appear in only one grism orientation). Thus, we fit the resulting NIRISS grism exposures and \hst\ photometry with models over a broad redshift range $z=[0,15]$, to allow for both low-$z$ and high-$z$ solutions. The best-fit model is considered the one with the highest $P(z)/\chi^{2}$ value. Associated 1D spectra are then optimally extracted from the data by \grizli, using the morphological model of the source as a reference for the position and extension (in the spatial direction) of the spectral trace.

\begin{figure*}
\center
 \includegraphics[width=0.85\textwidth]{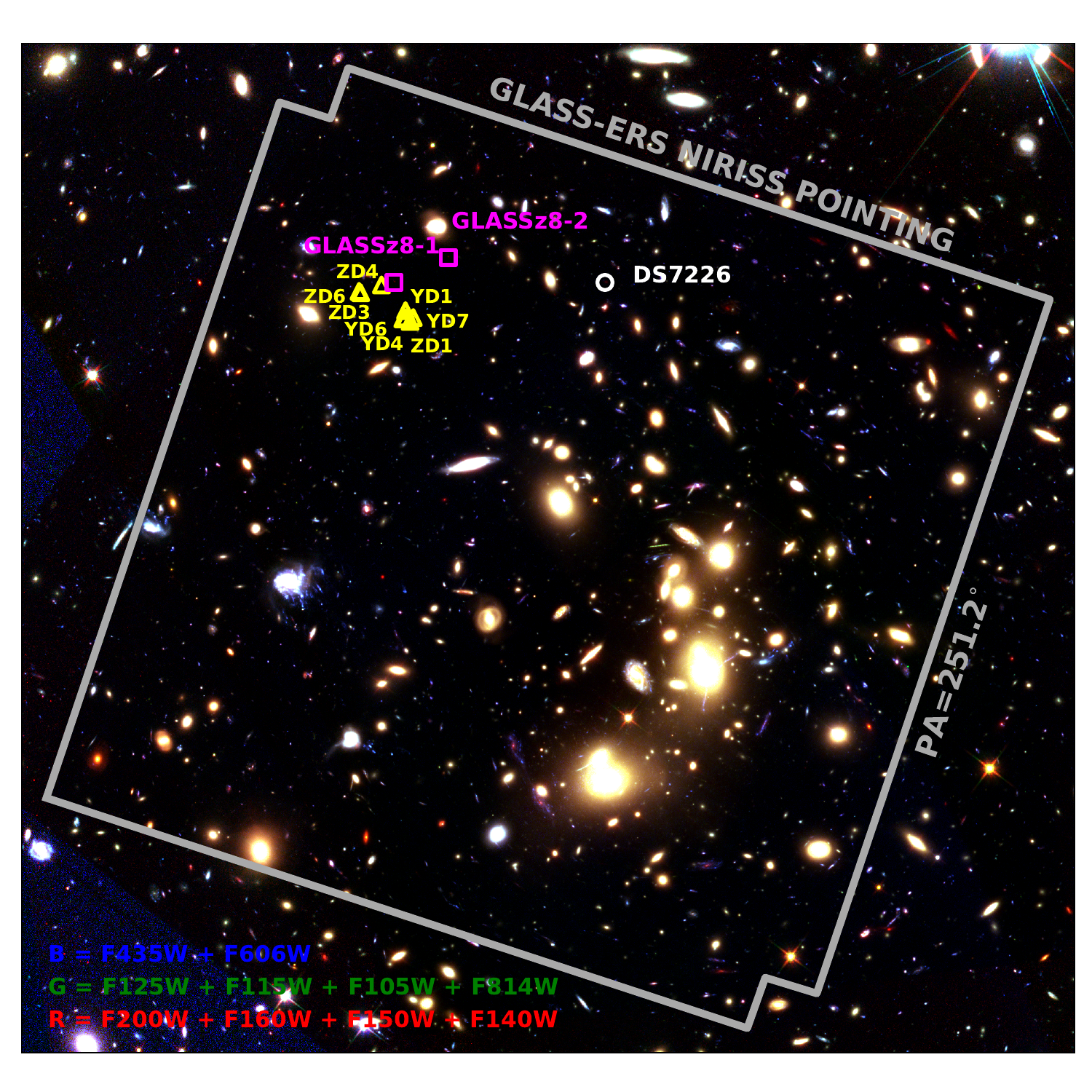}
 \caption{A false-color image of the Abell 2744 cluster using \textit{JWST}/NIRISS, \textit{HST}/WFC3 and \textit{HST}/ACS filters (red=F200W+F160W+F150W+F140W; green=F125W+F115W+F105W+F814W; blue=F606W+F435W), and the location of our 3 targeted $z\geqslant7$ candidate galaxies shown in magenta squares (if spectroscopically confirmed at $z\geqslant7$) or white circles (if confirmed as an low-$z$ interloper). The locations of the $z-$ and $Y-$dropout protocluster galaxies are marked in yellow triangles (adopting the \citealt{zheng14} nomenclature), for reference.
 }
 \label{fig:rgb}
\end{figure*}

\section{Spectroscopic Confirmation of $\lowercase{z}\geqslant7$ Galaxies With WFSS}
\label{sec:results}

\subsection{Target Selection}
Here we focus on the WFSS follow up of previously identified $z\geqslant7$ sources in the Abell 2744 cluster that (i) fall in the NIRISS FOV, and (ii) are sufficiently bright to be reliably detected via continuum measurements, with the aim of determining their spectroscopic redshifts. At $z\geqslant7$, the filters adopted by GLASS-ERS allow for spectral coverage of the Lyman-break out to $z\sim7.2-17.0$. Additionally, the filters also cover emission lines such as Ly$\alpha$ and \NV\ emission out to $z\sim17.0$, \CIV\ out to $z\sim13.5$, \HeII\ and \OIII] out to $z\sim12.5$, and \CIII] out to $z\sim11.0$.

We select our targets from a comprehensive compilation of photometrically-selected sources from \citet{zheng14}, \citet{zitrin14}, \citet{leung18}, \citet{ishigaki18}, \citet{bouwens22}, and the public \textit{Hubble} Frontier Fields (HFF) catalogs of \citet{shipley18} and \citet{castellano16}. Each of the compiled sources were originally selected as $z\geqslant7$ LBGs from NIR color cuts using \hst\ ACS+WFC3 data and photo-$z$ constraints. 

In this initial study, aimed at continuum detections, we limit our search to galaxies with a reported F160W magnitude (or F125W if F160W is not reported) of $m_{\rm AB}<26$ and quality flags indicating robust flux and/or photo-$z$ measurements. Such a choice is based on S/N predictions given by the ETC, indicating $>2\sigma$ detections in F115W (per pixel, assuming a $r=0.1''$ aperture) for sources with $m_{\rm cont}\sim26$ AB.

Accounting for overlap between studies, the compilation and cut comprises a total of 6 unique sources, 3 of which are sufficiently isolated from bright neighbours to be detected in our IR stack and segmentation map. The final sample of sources spans a range of observed F160W magnitudes $\sim24.6-25.6$ AB and photometric redshifts of $z_{\rm phot}\sim7.3-7.9$. The galaxies - referred to as GLASSz8-1, GLASSz8-2, and DS7226 - and their locations relative to the cluster center are shown in Figure \ref{fig:rgb}, which displays an \hst-NIRISS RGB image of the central cluster.

\subsection{Identifying Lyman Break Features}
Out of the 3 galaxy candidates compiled and identified here, we confirm 2 of them - GLASSz8-1 and GLASSz8-2 (formerly ZD2 and 2458 in \citealt{zheng14} and \citealt{castellano16}, respectively) to lie at $z\sim8$ based on a clear drop in flux blueward of (rest-frame) 1216 \AA\ indicating substantial absorption by intervening neutral hydrogen along the line of sight. Both galaxies are sufficiently isolated as to be clear of contamination from nearby objects. 

\begin{figure*}
\center
 \includegraphics[width=\textwidth]{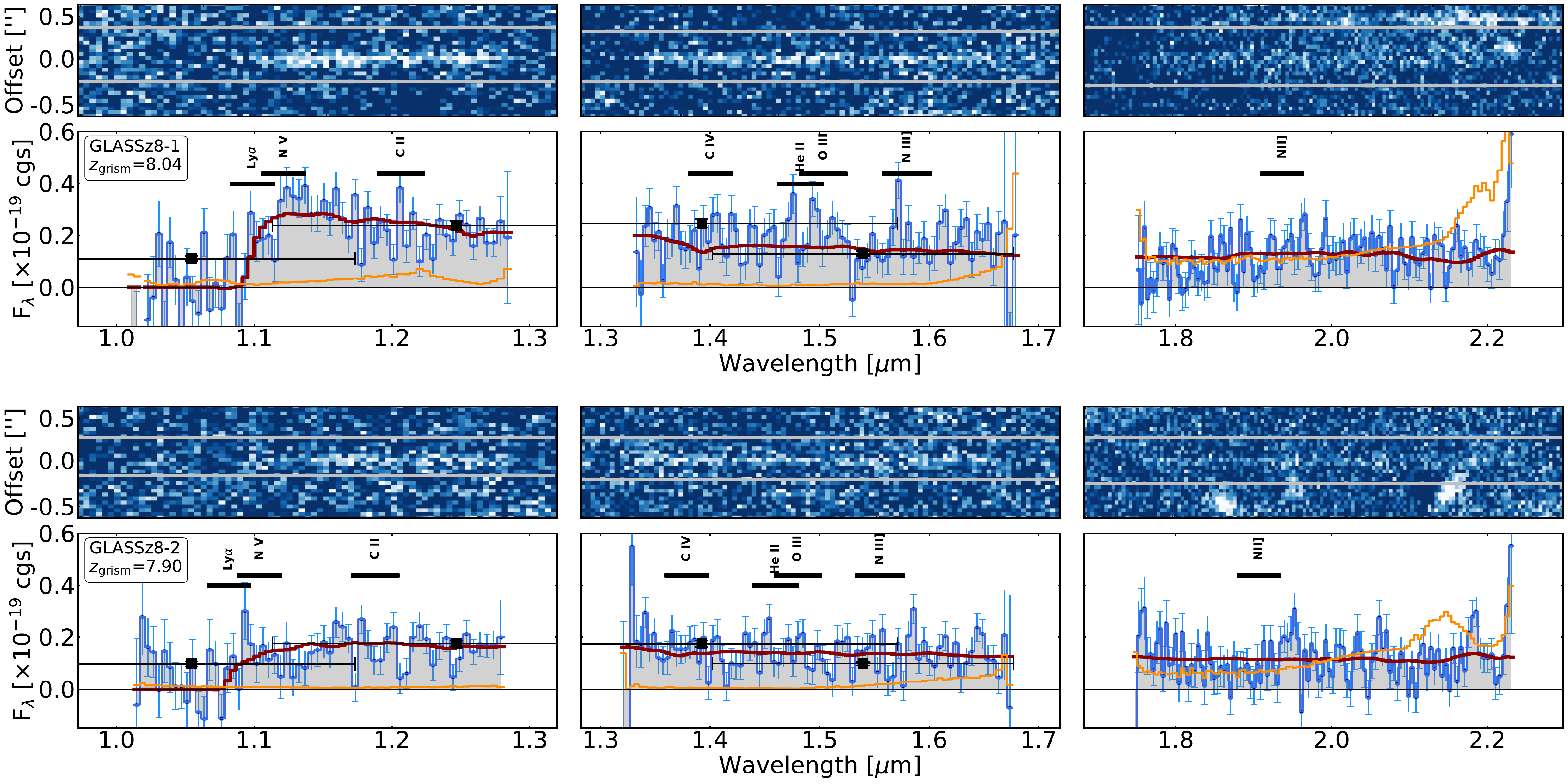}
 \caption{Postage stamps of confirmed $z>7$ LBGs in the Abell 2744 cluster field from NIRISS/WFSS observations. Each column shows results for a different filter (F115W, F150W, and F200W), while for each row top panels show the contamination-removed 2D drizzled stacks (over both grism orientations) as well as the spectral trace (grey lines) from which the 1D spectrum (blue lines and grey fill with associated scatter points and 1$\sigma$ uncertainties) is optimally extracted (bottom panel). We also show the best-fit \grizli\ continuum model in red, as well as the contamination model in orange, the latter of which is accounted for in the extracted spectrum. Additionally, the \hst\ photometry included in the fit are shown as black squares. We also indicate allowed wavelength windows for a suite of rest-frame UV emission lines, based on the spectroscopic redshift and associated $1\sigma$ uncertainty.}
 \label{fig:spectra}
\end{figure*}

We show their 2D spectra and 1D extracted fluxes in Figure~\ref{fig:spectra}, along with the best-fit \grizli\ model which places them at redshifts of $z_{\rm grism}=8.04\pm0.15$ and $z_{\rm grism}=7.90\pm0.13$, respectively. Redshift uncertainties are quoted as the $1\sigma$ standard deviation of the $P(z)$ at their $z\sim8$ locus. Inspecting Figure~\ref{fig:spectra}, we find the break is clearly identified in the F115W filter, where significant flux is detected redward of the break and only noise is visible blueward of it. GLASSz8-2 displays some apparently rising flux at the bluest end of the F115W. However, this is very close to the edge of the filter's sensitivity curve and the enlarged error bars make this consistent with noise from imperfect background subtraction. As expected for $z\geqslant7$ galaxies, the continuum is also detected (at reduced levels) across the entire wavelength ranges of the F150W and F200W filters, further supporting the confirmation. The reduced continuum levels at increasing wavelengths suggest blue UV slopes and absolute 1500 \AA\ magnitudes of $\beta=[-1.69_{-0.05}^{+0.04},-1.33_{-0.13}^{+0.14}]$ and $M_{\rm UV}=[-20.37_{-0.02}^{+0.02},-19.68_{-0.07}^{+0.05}]$ mag (for GLASSz8-1 and GLASSz8-2, respectively), corrected for magnification effects ($\mu\approx$2, see below) using the model of \citealt{bergamini22}), indicative of young, star-forming systems.

In both cases, we observe some excess flux at the expected position of Ly$\alpha$, however, such an excess is marginal and would require deeper and higher-resolution spectroscopy (e.g., with NIRSpec) to confirm. Subtracting the best-fit continuum model from the flux, normalizing by the continuum model, and integrating over the apparent line profile points to $3\sigma$ equivalent width upper limits of $<10-15$ \AA\ (rest-frame $<1.6$ \AA). We tabulate the galaxies' spectro-photometric properties in Table \ref{tab:table}, where we show that both spectroscopic redshifts are consistent with previously estimated photo-$z$'s. The remaining target displays a $z\sim2.6$ solution based on prominent emission line detections that classify it as a contaminant.


Both GLASSz8-1 and GLASSz8-2 are lensed by a factor of $\mu\approx2$ \citep{bergamini22} and are among the brightest of the target sample, with reported $H_{\rm 160}$ magnitudes of $m_{\rm AB}=25.56$ and $m_{\rm AB}=24.80$, respectively. We also find their positions on the plane of the sky place them within $\sim7-13.5$ arcsec from the $z\sim8$ overdensity, possibly indicative of some association between themselves and the other galaxies.

\begin{figure}
\center
 \includegraphics[width=\columnwidth]{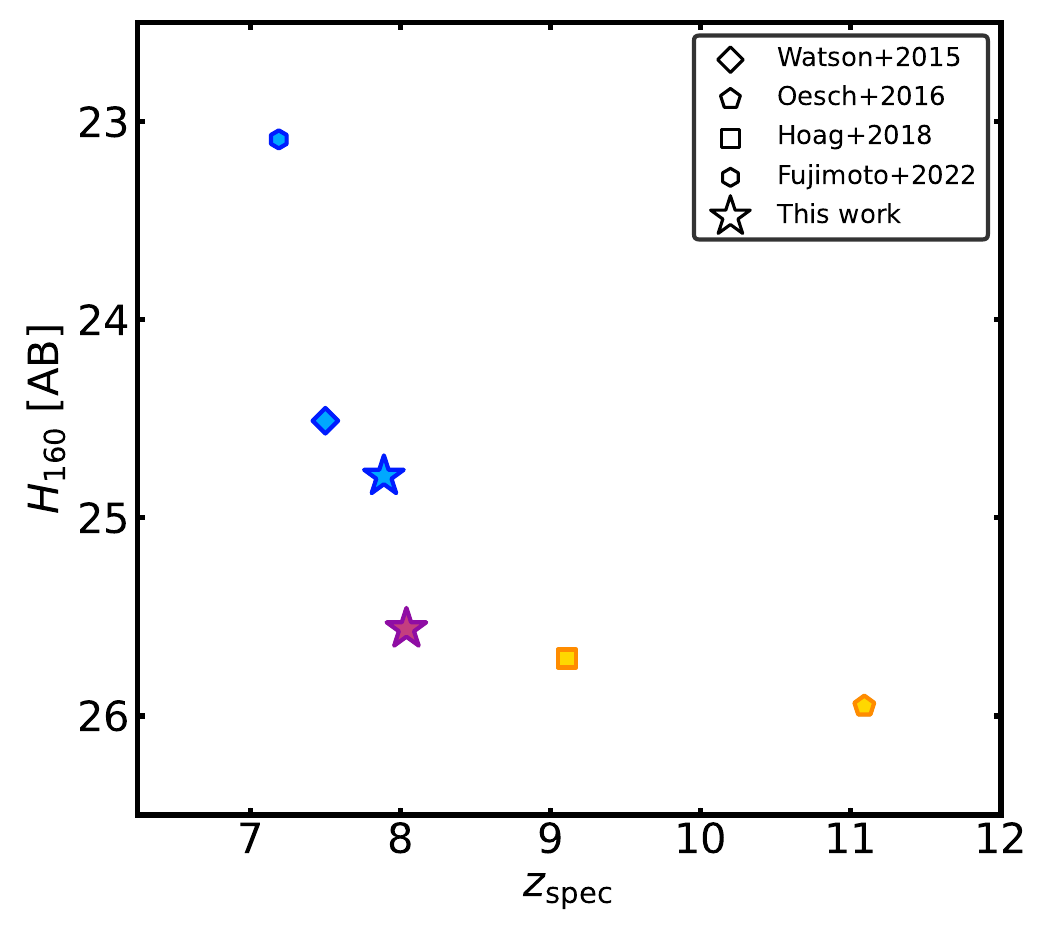}
 \caption{Redshifts and observed $H_{\rm 160}$ magnitudes from a compilation of confirmed LBGs in the literature \citep{watson15,oesch16,hoag18,fujimoto22} with clear continuum detections. Galaxy samples are color-coded according to their confirmed redshifts: $7\leqslant z<8$ are highlighted in blue, $8\leqslant z<9$ in purple, and $z>9$ in orange.}
 \label{fig:zgrim}
\end{figure}

The results shown here highlight the extraordinary potential of grism spectroscopy to determine spectroscopic redshifts independently of the Ly$\alpha$ emission seen in exceptional objects \citep[e.g.][]{finkelstein13,oesch15,zitrin15,rb16} and characterize unbiased samples of galaxies well into the Epoch of Reionization. To highlight this, in Figure \ref{fig:zgrim} we show a compilation of spectroscopically-verified LBGs at $z\gtrsim7$ with clear continuum breaks, irrespective of emission line detections. The number of galaxies unsurprisingly decreases as a function of redshift and remains exclusive to the apparently brighter sources -- the galaxies presented here increase the number of confirmed $z>7$ continuum detections in the literature by a factor of 1.5$\times$, thus serving as a powerful illustration of \jwst. Such samples will prove crucial to determine a representative picture of the sources that governed the reionization process and in this regard additional, pure-parallel grism observations over blank regions of the sky (e.g., GO~1571 PASSAGE, PI Malkan) will prove especially useful to obtain conclusions over the general galaxy population.

\subsection{Rest-Frame UV Line Emission}
In addition to the clear Lyman break, for GLASSz8-1 we also report tentative (i.e., $<5\sigma$) detections of rest-frame UV lines at expected wavelengths. The clearest identified lines are \HeII$\lambda$1640 \AA, the (unresolved) \OIII]$\lambda\lambda$1661,1666 \AA\ doublet, and the (unresolved) \NIII]$\lambda\lambda$1747,1749 doublet in the F150W filter. 
To quantify their statistical significance, we subtract the modelled continuum from the spectrum and measure the peak S/N ratio of each line from the residual, using again the error array from the flux-calibrated count-rate image as the uncertainty. The \HeII, \OIII], and \NIII] lines have peak S/N ratios of 2.7$\sigma$, 2.8$\sigma$, and 3.8$\sigma$, respectively. After verifying the Gaussian nature of the noise distribution, we find only 3 other pixels in the full F150W 1D spectrum have S/N$\gtrsim$2.75.
Although the lines are not especially strong, we note none of them appear close to any nearby residual contamination and their observed wavelengths and separations relative to each other are consistent with the interpretation. If confirmed at higher significance, the detections of \OIII] and \NIII] at such high redshift would prove a first.

The presence of rest-frame UV lines in $z>7$ galaxies is not unexpected, although have thus far been more prominently seen in strong Ly$\alpha$-emitters \citep[e.g.,][]{laporte17b,stark17,mainali18}.
Placing the line detections into context, the presence of \OIII] and \NIII] can be explained by both star-forming sources with low metallicities ($Z\sim0.001$) or non-thermal sources as they require moderate photon energies ($\sim$35 eV) to become ionized \citep{feltre16}. \HeII, in contrast, requires significantly higher energies of $\sim54$ eV from strong ionizing continuum photons suggestive of more extreme ionizing conditions (e.g., AGN or massive, low-metallicity stars). Low-metallicities systems are common among young, star-forming galaxies at $z\sim8$ \citep[e.g.,][]{strait20,rb21}, however only a handful of sources at $z>7$ display high-ionization lines similar to those reported here \citep[e.g.,][]{laporte17b,mainali18}. More detailed and higher S/N observations of such lines will be required to confirm their prevalence among the general galaxy population.


\section{Summary \& Conclusions}
\label{sec:conclusions}
In this Letter we present a first look at the \textit{JWST} NIRISS data from the GLASS-ERS survey taken over the Abell 2744 Frontier Fields cluster. We describe the procedures adopted by the survey to reduce, clean and model grism spectra. We showcase the power of \jwst\ by studying a sample of 3 $z\geqslant7$, $m_{\rm F150W}<26$ AB galaxy candidates with the aim of confirming their spectroscopic redshifts without relying on Ly$\alpha$. Our findings can be summarized as follows:

\begin{itemize}
    \item We confirm the spectroscopic redshifts of two out of three bright ($H_{\rm 160}\sim24.8-25.6$ AB) galaxies via the detection of strong Lyman breaks, placing them at redshifts of $z=8.04\pm0.15$ and $z=7.90\pm0.13$. The galaxies are located $\sim7-13.5$ arcsec from a $z\sim8$ protocluster, possibly indicative of some association between them. No strong and significant Ly$\alpha$ emission is detected in either of the galaxies. The third galaxy is found to be a $z=2.6$ interloper.
    
    \item In addition to the break, we find evidence for tentative ($\sim2.7-3.8\sigma$) detections of \HeII, \OIII], and \NIII] line emission, suggestive of a low-metallicity and star-forming system with potential non-thermal contributions.
\end{itemize}

The confirmation of galaxies at $z\sim8$ via their continuum is an important breakthrough in the study of cosmic reionization. Bypassing Ly$\alpha$ allows one to confirm and study galaxies in a way that is unbiased with respect to the ionization state of the surrounding cirgumgalactic and intergalactic hydrogen. Furthermore, and similarly to lower redshifts \citep{giavalisco02,shapley11}, we expect a large fraction of galaxies to not display strong emission lines. Those sources would have so far been missed and now become accessible with \textit{JWST}.

\begin{table*}
\centering
\begin{tabular}{lllccccccl}
\hline
GLASS ID & RA & DEC & $H_{\rm 160}$ & $\mu$ & $M_{\rm UV}/\mu$ & $\beta$ & $z_{\rm grism}$ & $z_{\rm phot}$ & Reference \\
         & [deg] & [deg] & [AB] & & [AB] & & & & \\
\hline
GLASSz8-1 & 3.60452 & -30.38047 & 25.56 & 2.00$^{+0.05}_{-0.04}$ & $-20.37_{-0.02}^{+0.02}$ & $-1.69_{-0.05}^{+0.04}$ & 8.04$\pm$0.15 & 7.90 & ZD2; \citet{zheng14} \\
GLASSz8-2 & 3.60135 & -30.37921 & 24.79 & 2.10$^{+0.06}_{-0.04}$ & $-19.68_{-0.07}^{+0.05}$ & $-1.33_{-0.13}^{+0.14}$ & 7.90$\pm$0.13 & 7.55 & 2458; \citet{castellano16} \\
\hline
\end{tabular}
\caption{A summary of NIRISS/WFSS spectroscopically-confirmed $z>7$ galaxies behind the Abell 2744 cluster, and their spectro-photometric properties. The last column lists the reference from which the photometric redshift and $H_{\rm 160}$ magnitude are taken. For magnification factors, we use the model of \citet{bergamini22}.}
\label{tab:table}
\end{table*}


\acknowledgments
This work is based on observations made with the NASA/ESA/CSA James Webb Space Telescope. The data were obtained from the Mikulski Archive for Space Telescopes at the Space Telescope Science Institute, which is operated by the Association of Universities for Research in Astronomy, Inc., under NASA contract NAS 5-03127 for JWST. These observations are associated with program JWST-ERS-1324. We acknowledge financial support from NASA through grant JWST-ERS-1324. KG and TN acknowledge support from Australian Research Council Laureate Fellowship FL180100060. CM acknowledges support by the VILLUM FONDEN under grant 37459. The Cosmic Dawn Center (DAWN) is funded by the Danish National Research Foundation under grant DNRF140. MB acknowledges support from the Slovenian national research agency ARRS through grant N1-0238. This research is supported in part by the Australian Research Council Centre of Excellence for All Sky Astrophysics in 3 Dimensions (ASTRO 3D), through project number CE170100013. We acknowledge financial support through grants PRIN-MIUR 2017WSCC32 and 2020SKSTHZ.

\bibliography{sample63}{}
\bibliographystyle{aasjournal}

\end{document}